\documentstyle[12pt,blois,psfig]{article}

\def\P3M{\hbox{$P^{3}M$}}

\def\AP3M{\hbox{$AdP^{3}M$}}

\def\cc2{c2}
\def\cc3{c3}
\def\cc4{c4}
\def\cc{c}

\def\AJ{AJ}

\def\ApJ{ApJ}

\def\MN{MNRAS}
\def\Nat{Nat}

\def\eg{{\it e.g.\ }}

\def\spose#1{\hbox to 0pt{#1\hss}}
\def\approxlt{\mathrel{\spose{\lower 3pt\hbox{$\sim$}}
	\raise 2.0pt\hbox{$<$}}}
\def\approxgt{\mathrel{\spose{\lower 3pt\hbox{$\sim$}}
	\raise 2.0pt\hbox{$>$}}}

\def\<{\thinspace}

\def\boxit#1{\vbox{\hrule\hbox{\vrule\kern3pt\vbox{\kern3pt
          #1 \kern3pt}\kern3pt\vrule}\hrule}}

\def\kpc{{\rm\thinspace kpc}}

\def\Mpc{{\rm\thinspace Mpc}}
\def\Msun{\hbox{$\rm\thinspace M_{\odot}$}}

\def\h50{\hbox{$\rm\thinspace h_{50}$}}
\def\h50m1{\hbox{$\rm\thinspace h_{50}^{-1}$}}

\begin{document}
\heading{COSMOLOGICAL GALAXY FORMATION}

\author{F. R. Pearce$^{1,3,}$, C. S. Frenk$^1$, A. Jenkins$^1$,
J. M. Colberg$^2$, P. A. Thomas$^3$, \\
H. M. P. Couchman$^4$, S. D. M. White$^2$, G. Efstathiou$^5$,
J. A. Peacock$^6$,\\
A. H. Nelson$^7$ (The Virgo Consortium)} 
{$^1$Physics Department, South Rd, Durham, DH1 3LE, UK}
{$^2$Max-Plank-Institut fur Astrophysik, 85740 Garching, Germany}
{$^3$CPES, University of Sussex, Falmer, Brighton, BN1 9QH, UK}
{$^4$Department of Physics and Astronomy, Univ. of Western Ontario, London, Ontario, N6A 3K7, Canada}
{$^5$Institute of Astronomy, Madingley Rd, Cambridge, CB2 OHA, UK}
{$^6$Physics Department, Edinburgh University, Blackford Hill, Edinburgh, EH9 3HJ, UK}
{$^7$Department of Physics and Astronomy, Cardiff University, Cardiff,
CF2 3YB, UK}
\begin{bloisabstract}

We discuss the first results from two successful simulations of
galaxy formation within a cosmological volume. With over
2000 large objects forming in each we have sufficient numbers to reliably
produce both galaxy correlation and luminosity functions. 
We find that the observed galaxy counts are well fitted by these models
and that the galaxies display an almost un-evolving correlation function
back to a redshift of 3 which closely resembles the featureless
observed form and amplitude. 

\end{bloisabstract}
\section{Introduction}

The quest for a detailed understanding of galaxy formation has become
one of the central goals of modern astrophysics.  On the observational
side data from the Keck and Hubble Space telescopes have
revolutionised our view of the high redshift Universe and there are
claims that the epoch of galaxy formation has already been observed
\cite{b97}. From the theoretical point of view the problem is
fundamental because its resolution involves the synthesis of work from
a wide range of specialities.  A full treatment requires consideration
of the early Universe processes that create primordial density
fluctuations, the microphysics and chemistry that precipitate star
formation within giant molecular clouds, the energy
exchange present in supernova feedback \cite{nw93}, 
the dissipational processes of cooling gas, the dynamics of
galaxies moving within a dense environment \cite{m96, fews}
and the large scale tidal torques which determine the
angular momentum profiles of the resultant objects.

Analytic approaches to the problem founder partly because of the lack
of symmetry. As high redshift observations show, real galaxies do not
form in a smooth, spherically symmetric fashion but rather form as
complicated collections of bright knots which merge and evolve into
normal galaxies \cite{sph95}.  Because of its
complexity this problem is often approached using numerical simulation.
In one of the first attempts \cite{esd}
demonstrated the a wide variety of observed galaxy properties
could be fitted by even the most basic of simulations. Unfortunately
their computational volume (like that of \cite{khw}) was too
small to produce a reliable correlation function and they had to 
stop the simulation at $z=1$ but this work clearly demonstrated the
potential for simulations of this kind. Both these groups used
smoothed particle hydrodynamics (SPH) to model the gas. In a 
complimentary investigation \cite{co} used a grid
based code, providing a useful cross-check of the results. More
recently several groups have attempted to tackle the problem
by coupling a semi-analytic approach to a collisionless simulation
and have again produced interesting results \cite{g97,k98}.

In this paper we simulate the process of galaxy formation within a
representative volume of the Universe in two contrasting cosmologies.
The volume is large enough -- $100\Mpc$ on a side -- that several
thousand galaxies form and we can reliably measure the galaxy
correlation function and study the effects of bias.  Previous work by
the Virgo Consortium \cite{j98} predicted the magnitude of
the bias that would be required to reconcile state-of-the-art
collisionless simulations to the observed galaxy correlation
function. Here we examine whether the inclusion of a gaseous component
and basic physics can indeed produce this bias.

\section{The simulation}

The simulations we have carried out are state-of-the-art SPH
calculations of 2 million gas plus 2 million dark matter particles in
a box of side $100\Mpc$ using a parallel adaptive particle-particle
particle-mesh plus SPH code \cite{pc}
We have completed both a standard Cold Dark Matter
(SCDM) run and a run with a positive cosmological constant ($\Lambda$CDM).
Both have the same parameters as detailed by \cite{j98}.
The baryon fraction was set from nucleosynthesis constraints,
$\Omega_bh^2=0.015$ and we assume an unevolving gas
metallicity of 0.5 times the solar value.  This leads to a gas mass
per particle of $2\times10^9\Msun$ for both runs.  As we typically
smooth over 32 SPH particles this gives us an effective minimum gas
mass resolution of $6.3\times10^{10}\Msun$.  We employ a comoving
Plummer gravitational softening of $10h^{-1}\kpc$ and the minimum SPH
resolution was set to match this.
 
\section{Underlying assumptions}

This simulation is based upon two fundamental assumptions. These
are that the feedback of energy due to supernovae explosions
can be approximated by assuming that this process effectively
imposes a mass resolution threshold. Objects below this
mass cannot form whereas objects above this mass are unaffected.
Comparing the simulated star formation rate to the observed
one shows that such an approximation does not lead
to a ridiculous star formation history. 
The second assumption is that once gas has cooled into tight,
dense blobs it is effectively decoupled from the surrounding
hot gas. This is equivalent to assuming that this gas has been
converted into stars or is isolated by additional physics
such as magnetic fields. 
The main further approximation we are forced to employ
for computational reasons is a spatial resolution of $10h^{-1}\kpc$. 
This is over twice the value we would have liked to have
used and leads to enhanced tidal disruption, drag and merging
within the larger clusters of objects.

The microphysics of star formation and supernovae explosions typically
have parsec lengthscales and so happen far below our resolution
limit. The combination of poor man's feedback and decoupling the
cold, dense gas effectively negates these effects (over which
we have no control anyway) for all objects above our resolution
threshold. All objects must cross this threshold at some time
(as a large object cannot simply {\it pop} into existence,
presumably via the merger of smaller, sub-resolution (and so
unresolved) fragments.
Our assumptions are equivalent to presuming that the absence of previous
levels of the hierarchy has no effect on the subsequent evolution
or properties of our objects and that these objects are large 
enough to resist the destructive effects of supernovae explosions.

It should be stressed that because of our relatively poor physical
resolution we can only predict accurate masses and positions for
our objects. We do not have sufficient resolution to resolve
internal structure and so cannot directly ascertain a morphological
type for our galaxies as each object typical only contains between
100 and 1000 particles.

\section{Analysis}

\begin{figure*} \centering \psfig{file=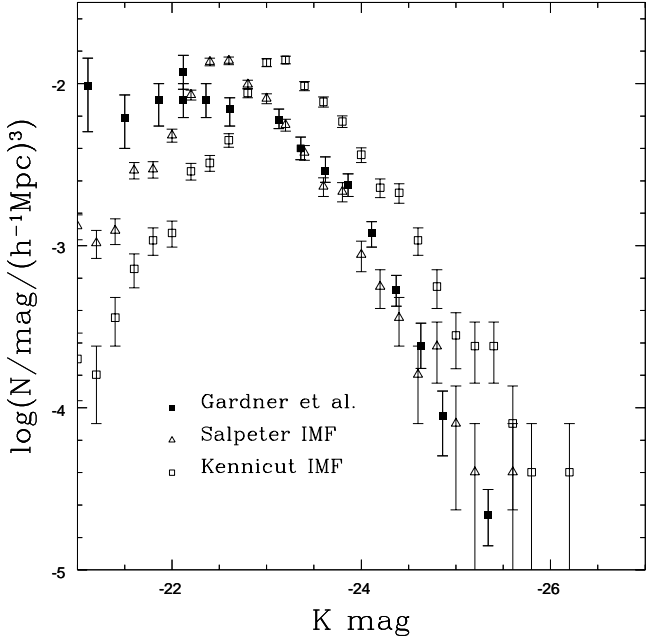,height=9cm}
 \caption{A comparison between the observed K-band luminosity function
and that obtained by the SCDM simulation with two different
assumptions for the stellar IMF.}
\end{figure*}

The presence of gas makes the identification of a set
of objects which we would like to equate to galaxies very
straightforward. Cooling leads to large collapse factors
and small knots of cold gas residing within a much hotter
halo. For the purposes of this paper we use a standard friends-of-friends
group finder with a small linking length to reliably extract
a group catalog from each simulation output time. 
At the end of the simulation 
there are over 2000 significant objects within each simulation
volume, roughly the observed number density.

The luminosity of any object is then extracted by tracking
each of the particles that finally reside inside each object
backwards in time and extracting the time at which each of them
first became cold and dense. Standard population synthesis
techniques (\eg \cite{bc}) can then be applied to produce 
the luminosity of each object in any desired pass band.

In Fig.~1 we show a comparison between the observed K-band luminosity
function \cite{gsfc97} and the SCDM simulation. The observations are
neatly bracketed by the simulation within the uncertainties imposed by
the choice of initial mass function. Clearly the {\it shape} of the
observations are well reproduced with no excess at the bright end.
The faint end slope is not well reproduced because of the relatively
high resolution imposed mass cut-off which was deliberately chosen to
be close to $L_*$. For the LCDM simulation the shape of the luminosity
function is also well modelled but the objects tend to be too
bright.  This is because the parameters chosen for this simulation
lead to a higher global fraction of cold, dense gas making all objects
more luminous. 

\section{Conclusions}

For the first time we have been able to model galaxy formation
within a large enough volume of the Universe that reliable 
calculations can be made of the galaxy correlation and luminosity
functions. As has been shown elsewhere in these proceedings
(\cite{jhere}) the observed correlation function is well fitted
by these models which also simultaneously fit the observed 
luminosity functions. The galaxies in our models are not only
reasonably distributed, they also have a sensible range of masses
and formation times.

\section*{Acknowledgments}

The work presented in this paper was carried out as part of the programme
of the Virgo Supercomputing Consortium using computers
based at the Computing Centre of the Max-Planck Society in Garching
and at the Edinburgh Parallel Computing Centre.

\end{document}